\documentclass[conference]{IEEEtran}
\IEEEoverridecommandlockouts
\usepackage{cite}
\usepackage{amsmath,amssymb,amsfonts}
\usepackage{algorithmic}
\usepackage{graphicx}
\usepackage{textcomp}
\usepackage{xcolor}
\usepackage{subfigure} 
\usepackage{hyperref}
\usepackage{adjustbox} % 加入 adjustbox 包
\definecolor{b2}{rgb}{0, 0, 0}
\definecolor{b}{rgb}{0, 0, 0}
\definecolor{b3}{rgb}{0, 0, 0}
\definecolor{b4}{rgb}{0, 0, 0}
\def\BibTeX{{\rm B\kern-.05em{\sc i\kern-.025em b}\kern-.08em
    T\kern-.1667em\lower.7ex\hbox{E}\kern-.125emX}}
\begin{document}

\title{Large Language Model (LLM)-enabled Graphs in Dynamic Networking

\author{Geng Sun,
        Yixian Wang,
        Dusit Niyato,~\IEEEmembership{Fellow,~IEEE},
        Jiacheng Wang,
        Xinying Wang,\\
        H. Vincent Poor,~\IEEEmembership{Life~Fellow,~IEEE},
        Khaled B. Letaief,~\IEEEmembership{Fellow,~IEEE}}
\thanks{
       G. Sun is with the College of Computer Science and Technology, Jilin University, Changchun 130012, China, and also with the College of Computing and Data Science, Nanyang Technological University, Singapore 639798 (email: sungeng@jlu.edu.cn).
       \par Y. Wang and and X. Wang are with the College of Computer Science and Technology, Jilin University, Changchun 130012, China (e-mail: yixian23@mails.jlu.edu.cn, wangxingying@jlu.edu.cn).
       \par D. Niyato and J. Wang are with the College of Computing and Data Science, Nanyang Technological University, Singapore 639798 (e-mail: jiacheng.wang@ntu.edu.sg, dniyato@ntu.edu.sg).
       \par H. V. Poor is with the Department of Electrical and Computer Engineering, Princeton University, Princeton, NJ08544, USA (e-mail: poor@princeton.edu).
       \par Khaled B. Letaief is with the Department of Electrical and Computer Engineering, Hong Kong University of Science and Technology, Hong Kong (e-mail: eekhaled@ust.hk).}
}
\maketitle

\begin{abstract}
Recent advances in generative artificial intelligence (AI), and particularly the integration of large language models (LLMs), have had considerable impact on multiple domains. Meanwhile, enhancing dynamic network performance is a crucial element in promoting technological advancement and meeting the growing demands of users in many applications areas involving networks. In this article, we explore an integration of LLMs and graphs in dynamic networks, focusing on potential applications and a practical study. Specifically, we first review essential technologies and applications of LLM-enabled graphs, followed by an exploration of their advantages in dynamic networking. Subsequently, we introduce and analyze LLM-enabled graphs and their applications in dynamic networks from the perspective of LLMs as different roles. On this basis, we propose a novel framework of LLM-enabled graphs for networking optimization, and then present a case study on UAV networking, concentrating on optimizing UAV trajectory and communication resource allocation to validate the effectiveness of the proposed framework. Finally, we outline several potential future extensions.
\end{abstract}


\section{Introduction}
\label{sec_Introduction}

\par As a basic framework for modeling complex relationships and structures, graphs play a vital role in \textcolor{b}{a number of} domains. In biology, graphs are used to \textcolor{b}{model} various complex networks within organisms, such as metabolic pathways, neural networks, or ecological relationships \textcolor{b}{among} species \cite{Pavlopoulos2018}. These graph models help scientists understand a variety of biological processes and interactions, leading to explore biodiversity and ecosystem stability. Furthermore, in social networks, graphs illustrate connections between individuals, revealing patterns of interaction, influence, and information flow \cite{Rezvanian2016}. By visualizing these networks, researchers can gain insights into social dynamics, behavior diffusion, and community structure. The versatile \textcolor{b}{applications} of graphs \textcolor{b}{highlight} their indispensable role in \textcolor{b}{understanding} complex phenomena and driving progress in different \textcolor{b}{fields}.

\par Traditional methods of graph analysis \textcolor{b2}{such as Transformers, BERT, and Graph Neural Networks (GNNs)} rely on predefined rules or algorithms and have achieved remarkable results in static data analysis \cite{Jin2023a}. However, when dealing with dynamic or evolving networks, they face the challenge of adapting to changing circumstances and integrating contextual information. In response to these limitations, there has been a growing interest in leveraging Large Language Models (LLMs) to enhance \textcolor{b4}{graph} analysis. By analyzing textual descriptions or annotations associated with nodes and edges in a graph, LLMs can reveal \textcolor{b}{explicit/implicit} relationships, infer missing information, and provide explanations for observed patterns. This synergy between graph-based representations and natural language understanding offers a powerful framework for gaining deeper insights into complex systems and addressing the shortcomings of traditional graph analysis methods \cite{Probierz2023}.

\par \textcolor{b}{LLM-enabled} graphs have demonstrated remarkable success in various domains. These achievements include enhancing semantic understanding, improving node classification, and predicting complex graphs \cite{Jin2023a}. \textcolor{b}{An} integration of LLMs into these applications demonstrates their potential to extract meaningful insights from complex data structures. Besides, by leveraging LLMs, especially its ability to capture complex patterns and dependencies in data, we can significantly enhance the understanding and analysis of dynamic networks. This integration is expected to lead to more accurate predictions of network behavior and improved real-time decision making. Therefore, \textcolor{b}{LLM-enabled} graphs undoubtedly have significant potential advantages, and this paper explores its application in dynamic networks. Overall, the main contributions of this paper are summarized as follows.

\begin{itemize}
    \item {We investigate and analyze \textcolor{b}{LLM-enabled} graphs from various aspects, including graphs \textcolor{b4}{to text and text to} \textcolor{b}{graphs. We also provide} the background and advantages of \textcolor{b}{LLM-enabled} graphs in dynamic \textcolor{b3}{networking. This is the first work that explores a novel applications of LLMs with graph for networking.}} 
    \item {We introduce \textcolor{b}{LLM-enabled} graphs and their applications in dynamic networks from the perspective of LLMs as different roles such as \textcolor{b4}{predictors, encoders and aligners}, including their principles, strengths, and \textcolor{b3}{limitations. They provide useful classification of LLMs' roles when applied to solve dynamic networking issues.}}
    \item{We propose a framework based on \textcolor{b}{LLM-enabled} graphs for networking optimization and verify its effectiveness through an example of the optimization of UAV trajectory and \textcolor{b}{communication resource} allocation in UAV \textcolor{b3}{networking. The framework is also general and can be applied to other types of networks.}}
\end{itemize}

\section{Overview of \textcolor{b}{LLM-enabled} Graphs}
\label{sec_overview}
\par In this section, we \textcolor{b4}{first give some background on graphs and then present related concepts and} applications of \textcolor{b}{LLM-enabled} graphs. Finally, we illustrate \textcolor{b4}{the potential for using LLM-enabled} graphs in dynamic networks.

{\color{b2}
\subsection{Graphs}
\label{sec_LLMs and Graphs}
\par Graphs \textcolor{b4}{are important data} structures used to represent a collection of entities and their relationships, including nodes and edges that connect \textcolor{b}{the} nodes. Moreover, graphs can be both a representation and a method for solving problems. For example, graphs are used to represent social interactions between users in social network analysis\footnote{\label{wiki}\url{https://www.ibm.com/docs/en/iii/9.0.0?topic=tool-social-network-graph}}, and knowledge graphs are widely employed in search engines and recommendation systems to support complex queries and reasoning by storing entities and their relationships \textcolor{b}{in} graph structures\footnote{\label{KG}\url{https://www.chatgptguide.ai/2024/02/26/what-is-knowledge-graph/}}. Furthermore, through \textcolor{b}{graph-based} algorithms such as shortest path algorithms (e.g., Dijkstra), network flow algorithms (e.g., Ford-Fulkerson), and graph traversal techniques (e.g., Depth-First Search or Breadth-First Search), we can analyze efficiency, robustness, and vulnerability to support decision making and network design. 

\par Recently, GNNs have \textcolor{b}{been used} in a variety of studies and applications of graphs. GNNs can extract and discover features and patterns in graph data, so as to meet the requirements of graph learning tasks such as clustering, classification, prediction, segmentation, and generation. For instance, GNNs can predict flow and congestion in transportation networks. The authors in \cite{LIU2023128525} developed a Dynamic Correlated Self-Attention (DCSA) module to capture dynamic node correlations and an Evolutionary Encoder-Decoder (EED) module to predict future traffic states.

\par However, processing large-scale graphs can be challenging due to increased computing and storage requirements, as well as the complexity in designing algorithms for tasks such as graph segmentation, search, and \textcolor{b}{path optimization}. Therefore, it is important to introduce LLMs to enhance the graphs.

\subsection{\textcolor{b}{LLM-enabled} Graphs}
\label{sec_Related Concepts}

\par LLMs are deep learning models based \textcolor{b4}{on the transformer} architecture, specifically designed for processing and generating human language. These models perform self-supervised learning on large text datasets to \textcolor{b}{understand} the complex structure and dependencies of the language. Based on unique language understanding and generation capabilities, LLMs are widely used in various language processing tasks such as text generation, language translation, sentiment analysis, and question-answering systems. 

\par An integration of LLMs with graphs demonstrates significant complementary advantages. Through graph embedding techniques, nodes and edges in a graph are transformed into vector representations, which can then be integrated with the word embeddings in LLMs to enhance the model's understanding of entities and relationships. \textcolor{b3}{For example, LLMs can analyze user community structures and predict the strength of relationships between users to provide more personalized social recommendations, thereby enhancing user experience.} Additionally, \textcolor{b}{the} introduction of GNNs, particularly in the joint optimization of pre-training and fine-tuning stages with LLMs, enables LLMs to handle \textcolor{b4}{text} and complex graphs. This deep integration \textcolor{b}{enables} the LLMs \textcolor{b}{for} multi-modal data processing. For example, in graph-to-text and text-to-graph applications, the LLMs can convert structural information from graphs into detailed textual descriptions, or conversely reconstruct textual descriptions into graphs.

\begin{figure*}[!hbt] 
	\centering
	\includegraphics[width =7in]{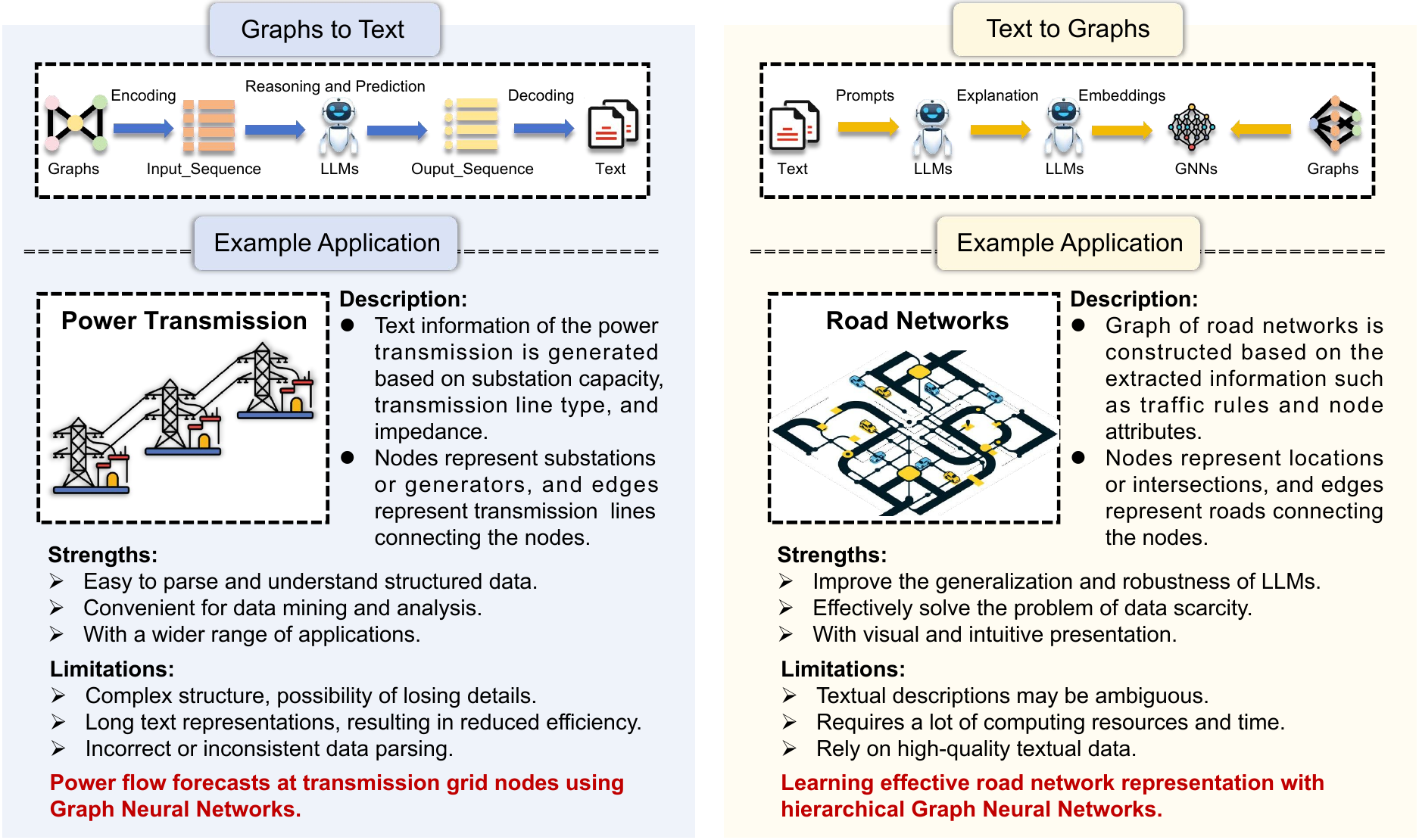}
	\caption{{The processes and applications of \textcolor{b}{LLM-enabled graphs:} from graphs \textcolor{b4}{to text and text to} graphs. LLMs play a crucial role in transforming graphical representations into textual information and converting \textcolor{b4}{text} back into graphs. This dual capability facilitates deeper analysis and understanding of complex systems, such as power transmission and road networks. }}
	\label{fig_LLMs1}
    \vspace{-1.2em}
\end{figure*}
\par Fig. \ref{fig_LLMs1} presents the process and application of \textcolor{b}{LLM-enabled} graphs from graphs \textcolor{b4}{to text and text to} graphs, and the details are as follows. 

\par \textbf{Graphs to \textcolor{b4}{Text:}} The non-hierarchical nature, collapse of remote dependencies, and structural diversity of graphs make it challenging to work directly with graph \textcolor{b4}{data. To deal with this issue it is helpful to convert graphs into textual form. With} this approach, we \textcolor{b}{can} better handle these complexities, simplify data representation, and leverage LLMs for analysis and processing, improving the interpretability and applicability of the data. For example, a power transmission graph can be regarded as a pure graph without textual or semantic information. The authors in \cite{Liu2023} used node lists and edge lists to represent the structure of the graph in natural language, where both nodes and edge lists are organized numerically and edges are divided in a sequential text format. The LLMs serve to accurately interpret the graph topology represented by these edges. \textcolor{b3}{For instance, they can identify critical nodes in a network where energy bottlenecks occur. By analyzing connectivity patterns and data flows, LLMs can also pinpoint potential performance degradation, thereby enhancing network optimization and reliability.}

\par \textbf{\textcolor{b4}{Text} to Graphs:} Converting \textcolor{b4}{text} into graphs helps us to extract information and present it visually. \textcolor{b}{Furthermore}, researchers can explore semantic relationships and complex interaction patterns \textcolor{b4}{between different texts by converting them to graphs,} thereby revealing \textcolor{b}{an} association between entities. This conversion provides a new perspective and method for deep understanding of texts. \textcolor{b3}{However, due to the nature of unstructured text (such as comments or social media posts), converting these texts into structured graphs poses a significant challenge. Therefore, it is essential to introduce LLMs.} Taking the road network application as an example, where nodes can be defined as intersections or locations of vehicles, and edges represent roads connecting these nodes\footnote{\label{chegg}{https://www.chegg.com/homework-help/questions-and-answers/road-network-road-network-modelled-graph-nodes-graph-represent-intersections-locations-nod-q98363392}}. This detailed definition of structure enables LLMs to understand the relationships between nodes and edges, and generate relevant text vectors, which are then combined with GNNs to present a road network graph, thus enhancing the visualization and understandability of the graph information \cite{He2023}.
%For instance, molecules are represented as graphs, where nodes represent atoms, edges represent chemical bonds, and some text represents related properties. Therefore, LLMs can be trained based on the molecules to produce the novel molecules. The authors in \cite{Skinnider2024} uniquely identify a molecule with the string SMILES or SELFIES, and discover that introducing invalid outputs can actually enhance the performance of LLMs. 

{\color{b2}
\subsection{Applications of \textcolor{b}{LLM-enabled} Graphs}
\label{sec_Literature}

\par \textbf{Molecule Modeling and Analysis:} Molecules are usually represented by graphs and paired with text of their basic properties. Joint modeling of both the molecular structure and the associated rich knowledge is important for deeper molecule understanding. For instance, the authors in \cite{Jablonka2024} fine-tuned LLMs (GPT-3) to answer chemical questions in natural language with the correct answer. Specifically, they study classification tasks (e.g., transition wavelength of 2-phenyldiazenylaniline is categorized into ``high" or ``low"), regression tasks to predict the value of a chemical property (i.e., floating-point numbers), and inverse design tasks (i.e., \textcolor{b}{molecules, whose structure can be represented as a graph).} 

\par \textbf{E-Commerce:} \textcolor{b4}{LLMs use language} understanding and presentation learning abilities to optimize \textcolor{b}{product advertisement and recommendation} to \textcolor{b}{improve} user satisfaction. This is essential to promote the development of e-commerce platforms and enhance user loyalty.} For example, the authors in \cite{Yang2021} introduced GraphFormers, an integration of LLMs and GNNs for understanding text graphs better. Through the progressive learning strategy, the framework makes full use of the neighborhood information in the linked text and enhances the ability of the model to integrate the graph information, thus improving the representation quality and user satisfaction.

\subsection{Lessons Learned}
\par From the applications, we can summarize several key advantages of \textcolor{b}{LLM-enabled} graphs.
\begin{itemize}
\item {\textbf{Enhanced Data Interpretation:} LLMs can automatically identify implicit relationships and entity properties within texts, accurately annotating nodes and edges in the generated graph.}

\item{ \textbf{Improved Semantic Representation:} LLMs are capable of understanding and \textcolor{b}{modeling} complexity of real information, and they provide rich and multi-dimensional semantic layers for graphs.}

\item{ \textbf{Flexibility and Generalization:} LLMs have the ability to extract information from both structured and unstructured data simultaneously. Moreover, LLMs can effectively process data from existing datasets and generalize novel information to promote research and innovation.}
\end{itemize}

\begin{figure*}[!hbt] 
	\centering
	\includegraphics[width =7in]{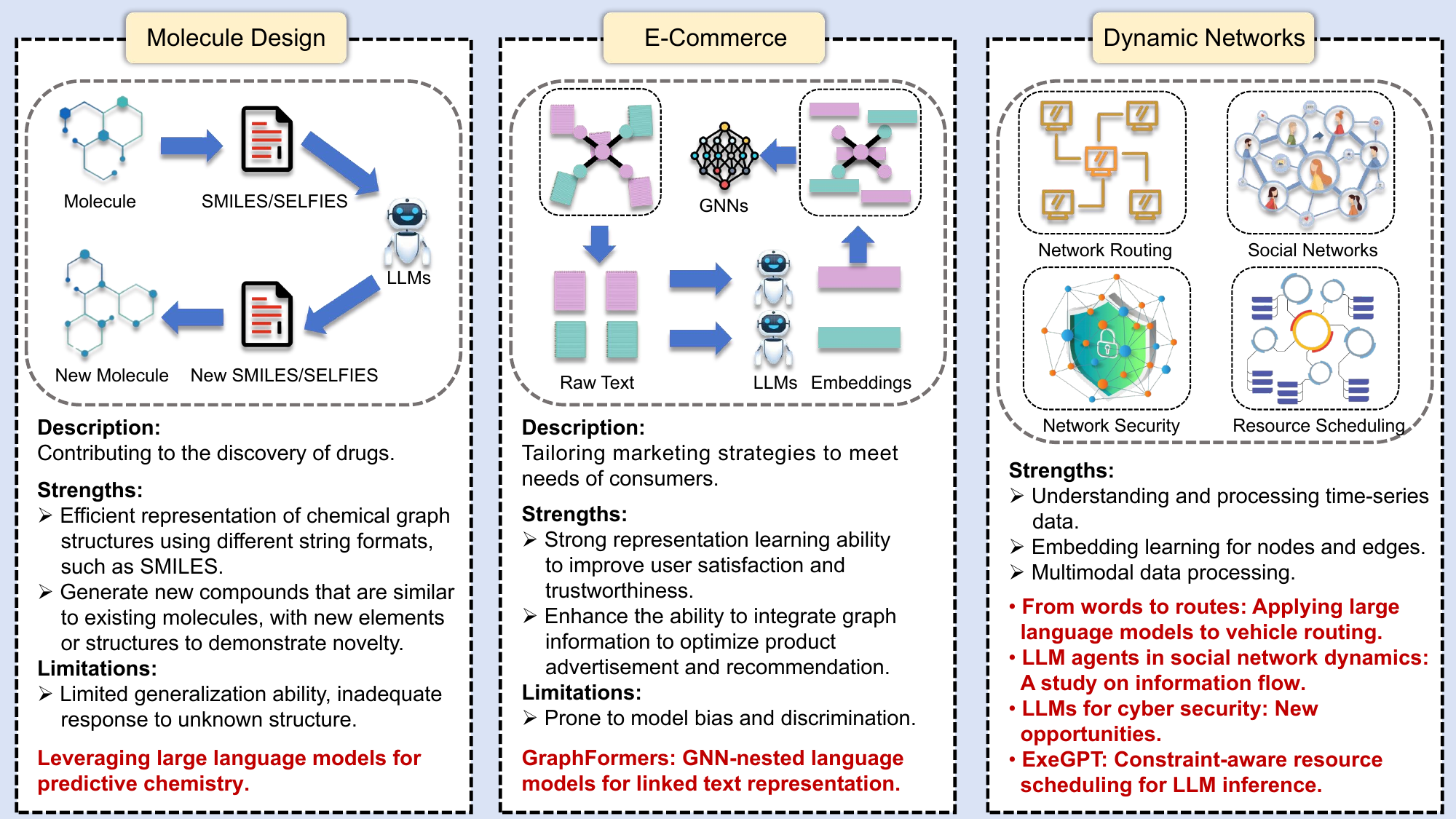}
	\caption{The summary of \textcolor{b}{LLM-enabled} graphs in different domains and dynamic networks. The \textcolor{b}{applications} of LLMs technology \textcolor{b}{have} led \textcolor{b4}{to important results} in drug discovery of molecule design, and personalized marketing of E-commerce. Besides, LLMs \textcolor{b4}{have significant potential} in dynamic networks and are expected to bring significant improvements to dynamic network application scenarios.}
	\label{fig_LLMs2}
 	\vspace{-1.5em}
\end{figure*}

\par From these applications and advantages, \textcolor{b}{LLM-enabled} graphs have demonstrated strong problem understanding and processing capabilities in the abovementioned domains. However, in dynamic networking, existing studies mainly rely on conventional GNNs. Despite impressive computing efficiency, they still face \textcolor{b}{certain} limitations. For instance, the structure and characteristics of dynamic networks are constantly changing, requiring frequent model updates to adapt to new data, which can lead to inefficient model training and inference. \textcolor{b}{Moreover}, it is difficult to capture time series \textcolor{b}{dependencies, e.g., packet flows,} in the graph and integrate information on multiple time scales. Therefore, it \textcolor{b4}{is of interest to} introduce \textcolor{b}{LLM-enabled} graphs to further enhance the capabilities in various aspects of dynamic networking. The advantages of this are as follows.
\begin{itemize}
\item{\textbf{Understanding and Processing Time-Series Data:} LLMs are capable of handling long-term dependencies and time-series data, making them highly suitable for analyzing patterns and trends in dynamic networks.}

\item{\textbf{Embedding Learning for Nodes and Edges:} By generating powerful embedding vectors, LLMs can capture the complex characteristics of entities and relationships to better represent dynamic nature of networks.}

\item{\textbf{Multi-modal Data Processing:} The multi-modal processing of LLMs integrate different data types (such as text, images, channel measurement, network congestion, and other types of sensing data) into dynamic networks, providing a richer and more nuanced perspective for analyzing complex networks and interactions.}
\end{itemize}

\par Fig. \ref{fig_LLMs2} provides a summary of \textcolor{b}{LLM-enabled} graphs. Clearly, \textcolor{b}{LLM-enabled} graphs have their unique advantages and challenges in different domains, and it is necessary to explore its applications in dynamic networks.

\section{\textcolor{b}{LLM-enabled} Graphs in Dynamic Networking}
\label{sec_LLMs-enabled Graphs in Dynamic Networking}
\par This section \textcolor{b4}{introduces} \textcolor{b}{LLM-enabled} graphs and their applications in dynamic networks from the perspective of LLMs as \textcolor{b4}{predictors, encoders and aligners, including} basic principles, strengths, and limitations.

{\color{b2}
\subsection{LLMs as \textcolor{b4}{Predictors}}
\par For graph-related prediction tasks such as classification and reasoning, LLMs mainly solve the problem of non-sequential and structural complexity of graphs by converting the graph structure into a serialized format that can be processed by the model. For example, converting graphs into sequences (such as graph descriptions or hidden representations generated by GNNs) for the LLMs to process directly. In \cite{Zhao2023}, the authors proposed a framework called GraphText that generates a graphic text sequence by traversing the syntax tree derived from the graph. \textcolor{b}{Specifically,} the syntax tree encapsulates node attributes and inter-node relationships such as those in a social network, and \textcolor{b4}{LLMs are used} to process this sequence. 

\begin{figure*}[!hbt] 
	\centering
	\includegraphics[width =7in]{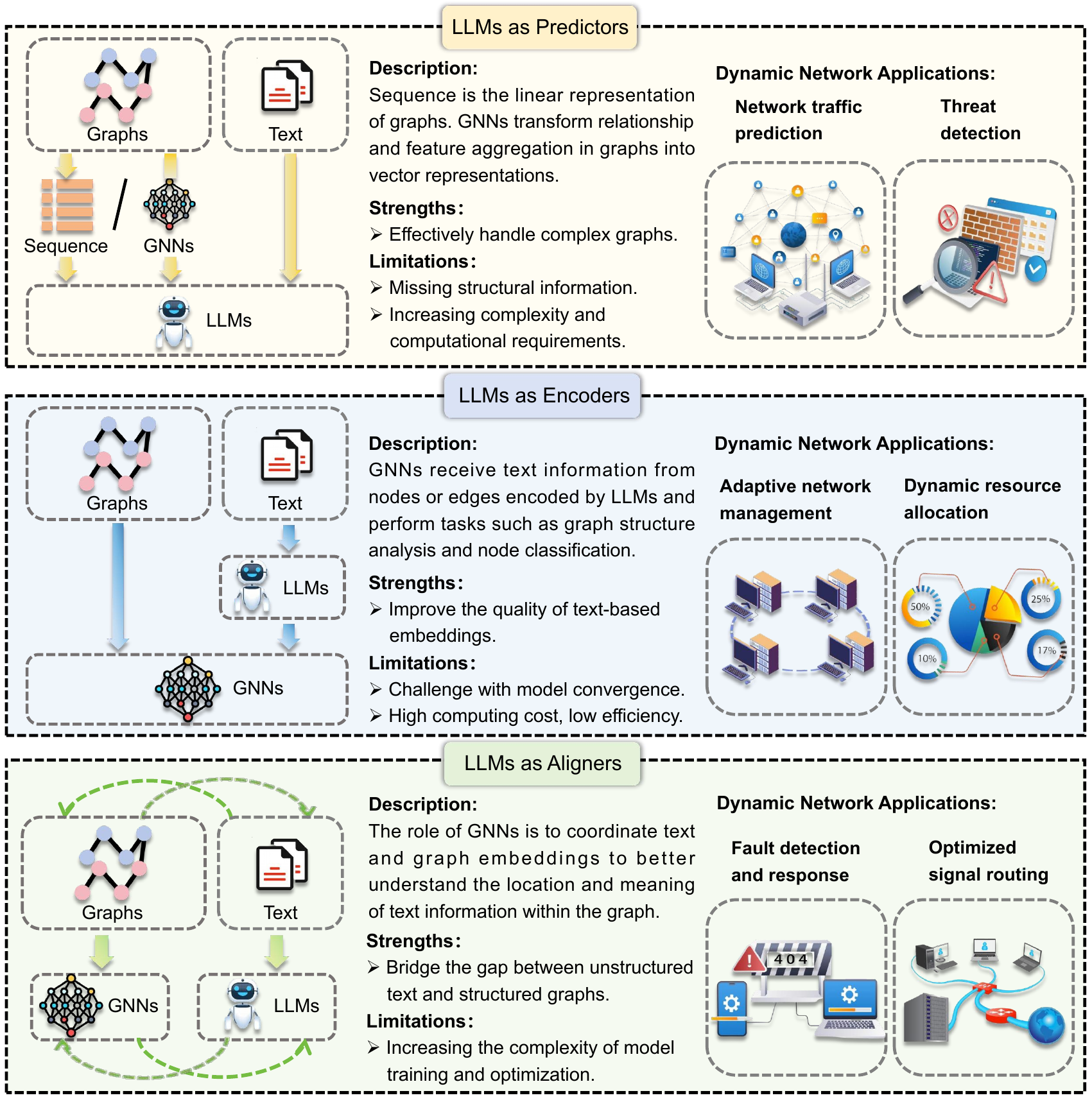}
	\caption{The summary of \textcolor{b}{LLM-enabled} graphs and their applications in dynamic networks \textcolor{b4}{from the perspectives of using LLMs as predictors, encoders and aligners.} \textcolor{b4}{LLMs as predictors can handle} serialized graph structures or vectors generated by GNNs. As \textcolor{b4}{encoders}, LLMs transform text information of nodes \textcolor{b}{and} edges into vector representations, enriching GNNs with textual data representations. Additionally, LLMs \textcolor{b4}{as aligners generate text} embeddings that coordinate with the structural embeddings produced by GNNs, facilitating alignment between \textcolor{b4}{text} and graphs.}
	\label{fig_LLMs3}
  	\vspace{-1.2em}
\end{figure*}

\par These methods can \textcolor{b}{capitalize on} the ability of LLMs to analyze and interpret complex graph patterns. For instance, in data communication networks, \textcolor{b4}{LLMs can be used} to analyze unstructured data from various sources, including Internet of Things (IoT) device logs and network traffic, aiding in better network traffic management and threat detection. In \cite{ALWAHEDI2024167}, LLMs can parse through gigabytes of log data from a smart home system to identify unusual patterns that might indicate a cyberattack, such as a sudden spike in outbound data suggesting data exfiltration or other malicious activities, thereby enhancing the security of IoT systems. However, there are some challenges, such as converting to sequences may lose structural information, and modifying the \textcolor{b4}{LLM} architecture may increase complexity and computational requirements in communication networks.

\subsection{LLMs as \textcolor{b4}{Encoders}} 
\par \textcolor{b4}{Using an LLM as an encoder} enhances graph analysis by encoding text information from nodes or edges into feature vectors. These vectors become the initial inputs for GNNs, providing richer textual data representations. For example, the authors in \cite{Chen2023} proposed a label-free node classification method called LLM-GNN. This method utilizes pseudo-labels generated by LLMs to expand the size of labeled data, thus reducing the reliance on actual labeled samples. Subsequently, it combines these pseudo-labels with a small number of real labeled samples to perform node classification tasks using GNNs for learning and prediction. 

\par The primary advantage of these methods is the improved quality of text-based embeddings. LLMs can understand and encode complex relationships within text, resulting in more informative embeddings that help GNNs understand the structure of the graph in a semantically rich way. For instance, in wireless networks, LLMs provide powerful support for intelligent and adaptive network management. In \cite{Zou2023}, the authors explored the use of LLMs to achieve collaboration among multiple wireless generating agents. Each agent \textcolor{b4}{employs an LLM as an encoder} to transform information perceived from its environment and domain-specific knowledge acquired from the cloud or other devices into vectors. These agents then share knowledge and state information by exchanging the encoded information via wireless communication, enabling collaborative planning and execution of complex tasks. \textcolor{b3}{This approach enhances network intelligence and automation, while also optimizing resource utilization and energy efficiency to improve wireless network performance.} \textcolor{b4}{However, using LLMs as encoders presents} challenges in terms of model convergence, especially the complexity introduced \textcolor{b4}{by integrating LLMs with GNNs. In} addition, the computational cost of LLMs can be very high, making the entire process less efficient for large-scale applications. 
 
\subsection{LLMs as \textcolor{b4}{Aligners}}
\par \textcolor{b4}{For using LLMs as aligners, their} natural language understanding ability is used to generate text embeddings, which are coordinated with the structural embeddings of GNNs to achieve text and graph alignment. For instance, GLEM was proposed \cite{Zhao2023a}, which integrates graph structure and language learning within a variational Expectation-Maximization (EM) framework to address large-scale text-attributed \textcolor{b4}{graph} learning problems. The two modules of E-step and M-step are alternately updated in this framework so that they influence and promote each other. Among them, the E-step is used to optimize the LLMs, while the M-step is used to train the GNNs.

\par These methods bridge the gap between unstructured text and structured \textcolor{b4}{graphs}, leading to deeper insights and better model performance in tasks involving complex relationships, especially in \textcolor{b}{wireless network applications.} In terms of signal transmission, GLEM utilizes the temporal attributes of TAGs to capture the dynamics of signal propagation in the network. By updating node representations at each time step, GLEM tracks the paths of signal propagation and optimizes them based on both historical data and real-time information, enabling faster and more reliable signal \textcolor{b4}{propagation. Moreover, taking a wireless} network management system as an example, the collaboration between LLMs and GNNs facilitates efficient fault detection and resolution. LLMs analyzes fault reports (e.g., ``Bandwidth is low on router X") and translates them into high-dimensional embeddings that capture the semantics of network problems. GNNs process the graph structure of the network, including nodes (e.g., routers or switches) and connections to create graph embeddings. These embeddings are then aligned, adjusting the text embeddings to precisely match the corresponding graph embeddings \cite{li2024survey}, \textcolor{b4}{and thereby enabling} accurate \textcolor{b}{management} and troubleshooting in wireless \textcolor{b}{networks.} However, maintaining consistency between different data types may increase the complexity of model training and optimization.
}

\par Fig. \ref{fig_LLMs3} provides a summary of the \textcolor{b}{LLM-enabled} graphs and their applications in dynamic networks from the perspective of LLMs as \textcolor{b4}{predictors, coders and aligners. Here,} different roles LLMs have their unique advantages and challenges.

\begin{figure*}[!hbt] 
	\centering
	\includegraphics[width =7in]{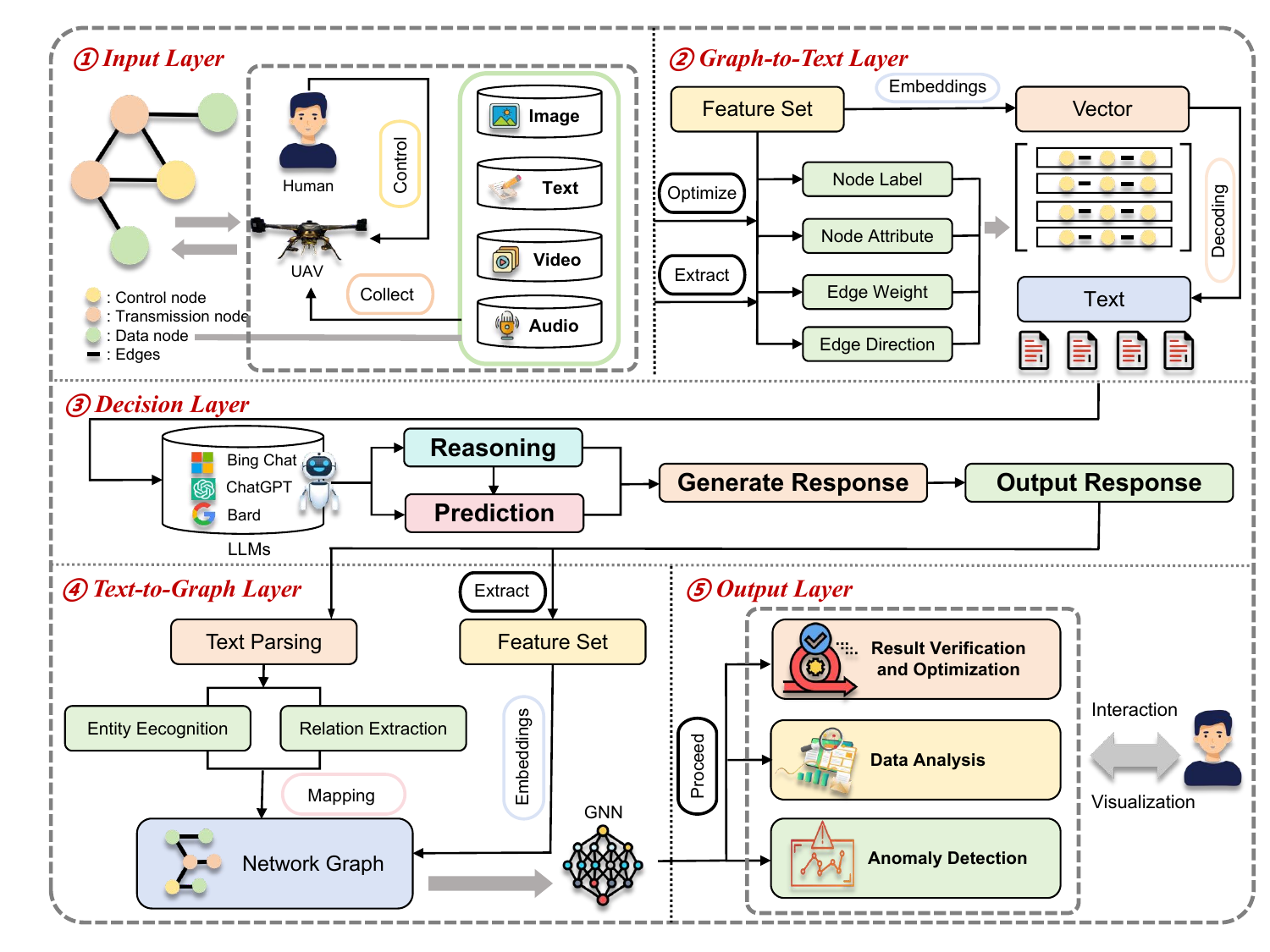}
	\caption{{The structure of the proposed \textcolor{b}{LLM-enabled} graphs framework. The framework is based on a layered architecture consisting of an input layer, a graph-to-text layer, a decision layer, a text-to-graph layer, and an output layer. The input layer receives requests related to a dynamic network graph. The graph-to-text layer employs prompt engineering to extract features from requests, and then converts them to \textcolor{b4}{text} via embeddings. The decision layer utilizes a pluggable LLM to generate responses. The text-to-graph layer extracts features from \textcolor{b4}{the text generated by the LLM to} construct the graph, which is then processed \textcolor{b4}{by a GNN}. The output layer analyzes the generated results and interacts with the user.}}
	\label{fig_LLMs5}
  	\vspace{-1.2em}
\end{figure*}

\section{\textcolor{b}{LLM-enabled} Graphs Framework for Dynamic Networking Optimization}
\label{sec_Optimzation}
\par In the section, we propose an \textcolor{b}{LLM-enabled} graphs framework for networking optimization. Then, \textcolor{b4}{considering the specific problem of UAV} networking, \textcolor{b4}{we study the} optimization of UAV trajectory and \textcolor{b}{communication resource} allocation to demonstrate the effectiveness of the proposed framework.
\subsection{Motivation and Challenges}
\label{sec_Motivation and Challenges}
\par LLMs provide a comprehensive toolkit for dynamic network analysis and optimization, including serialization of graph structures and encoding of textual data from network nodes and edges into feature vectors. Thus, this integration of LLMs into dynamic network graphs facilitates detailed and adaptive network designs. However, there are still various challenges in applying \textcolor{b}{LLM-enabled} graphs for dynamic networks.

\begin{itemize}
\item {\textbf{Maintenance of Real-Time Data:} In dynamic networks, the update and integrity of real-time data are critical for \textcolor{b}{LLM-enabled} graphs to make accurate predictions. Incomplete or outdated data can compromise effectiveness of decision-making and strategy optimization.}

\item{\textbf{Resource Management and Optimization:} The continuous change of network structures and fluctuations of data pose challenges to the demand and utilization of resources. LLMs need to adjust resource allocation in real-time to ensure timely data processing and transmission.}

\item{\textbf{Adaptation and Stability:} LLMs need to be adaptive and robust to handle changing behaviors and states. The self-tuning capability optimizes performance by adjusting parameters in real-time, while robustness ensures quick recovery from abnormal situations \textcolor{b}{such as} network \textcolor{b}{failures or cyber attacks.}}
\end{itemize}

\par \textcolor{b}{In} dynamic networks, LLMs face complex challenges in graph analysis that require innovative solutions. Inspired by the remarkable capabilities of LLMs, we propose an \textcolor{b}{LLM-enabled} framework to address these challenges.
%LLMs exhibit outstanding advantages in graph analysis for dynamic networks, but they also face complex challenges that require innovative solutions. Inspired by the remarkable capabilities of LLMs, we propose an LLMs-enabled graphs framework to address these challenges.

\subsection{Proposed Framework}
\label{sec_Proposed Framework}
\par As shown in Fig. \ref{fig_LLMs5}, our proposed framework \footnote{\label{our}\url{https://yx2024.github.io/}} follows a layered architecture, which consists of five \textcolor{b}{layers, i.e., the} input layer, graph-to-text layer, decision layer, text-to-graph layer, and output layer}.
%\par In this subsection, we introduce the proposed framework. First, we convert the problem in dynamic networks that can be represented by graphs into the text form. Next, we process and analyze these texts using LLMs to generate new texts output. Finally, we convert the generated texts back into graphs.
\begin{itemize}
\item {\textbf{Input Layer:} The input layer receives requests that can represent information related to a graph in a dynamic network, including a control node (e.g., a UAV operator), transmission nodes (e.g., UAVs), and data nodes that need to be processed by UAVs, allowing multi-modal data as input, such as text, images, video, and audio.}

\item{ \textbf{Graph-to-Text Layer:} We first use prompt engineering to optimize the graph information received \textcolor{b4}{from the input} layer and extract the most representative and information-rich features. Then, the graph information is transformed into vector representations \textcolor{b4}{that the system} can understand and process through embeddings to facilitate downstream model processing such as text generation or graph classification models. Finally, numerical features in the vector are decoded into understandable semantic features to generate the corresponding text description.}

\item{ \textbf{Decision Layer:} The decision layer \textcolor{b4}{employs a suitable LLM such} as Bing Chat, Chatgpt4, \textcolor{b4}{or Bard} to make decisions. \textcolor{b4}{Specifically, the LLM analyzes} and processes the text from the graph to text layer, combines the contextual information of the input text for inference and prediction, and generates new text as output to support subsequent task execution.}

\item{ \textbf{Text-to-Graph Layer:} In the text-to-graph layer, the key features of nodes and edges are extracted from \textcolor{b4}{the text generated by the LLM to} construct the graph structure. Then, the graph structure is processed \textcolor{b4}{by a GNN, and} the information of nodes, edges and text attributes are embedded. Finally, graph and text embeddings are merged to form an integrated vector representation for downstream tasks.}

\item{ \textbf{Output Layer:} The output layer will verify, analyze, monitor and present the results of dynamic network graphs, and interact with users, making the complex graph data easier to understand and utilize.}
\end{itemize}

\subsection{Case Study: }
\label{sec_Case}
\begin{figure*}[!hbt] 
	\centering
	\includegraphics[width =7in]{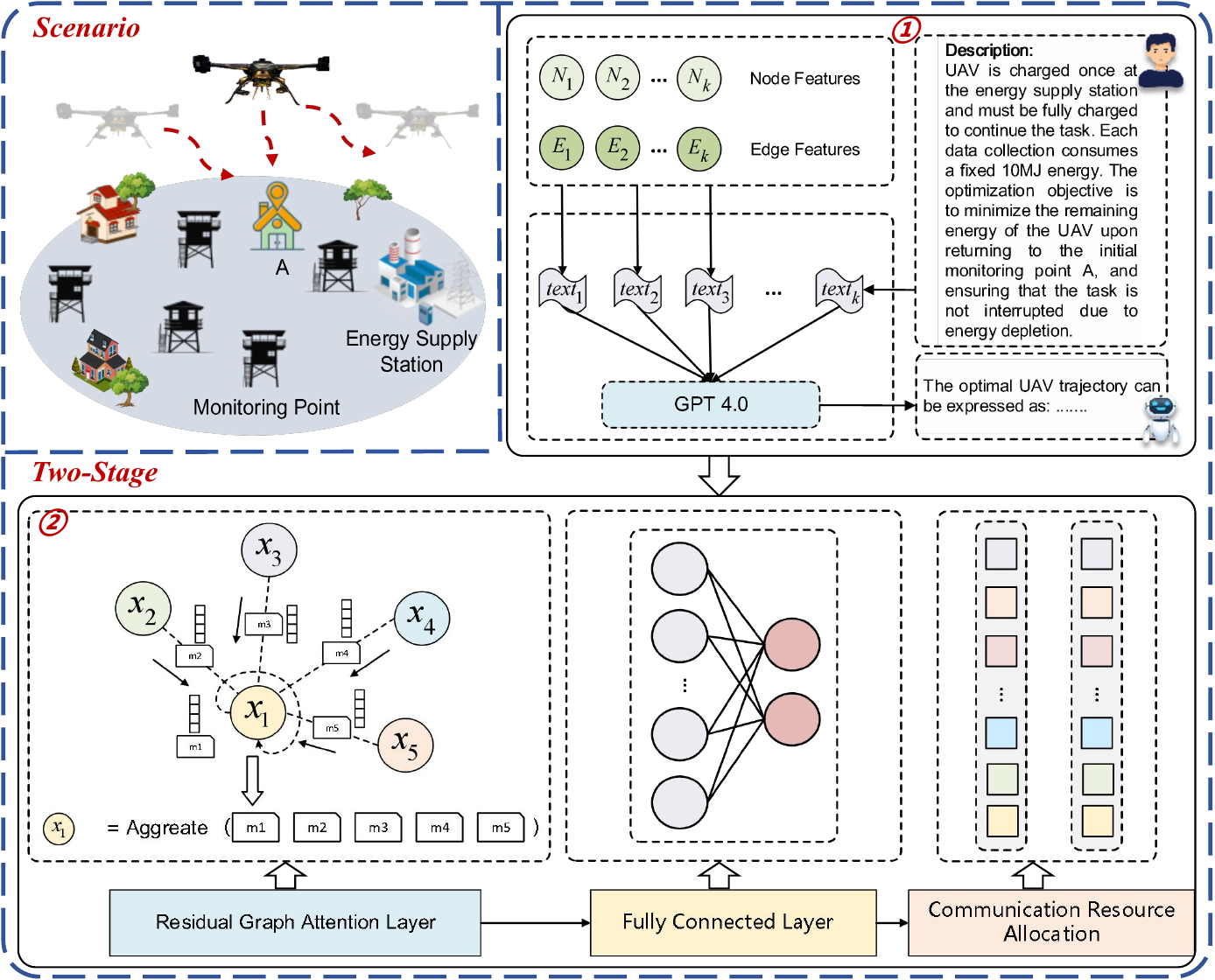}
	\caption{{The scenario module presents a system model of \textcolor{b4}{the considered scenario}. The two-stage joint optimization strategy includes the first stage module and the second stage module. \textcolor{b4}{Specifically, an LLM is used to generate a UAV} trajectory in the first stage, followed by \textcolor{b4}{employing a GNN} in the second stage for \textcolor{b}{communication resource} allocation. The three modules interact with each other to \textcolor{b4}{obtain a UAV} trajectory and \textcolor{b}{communication resource} allocation that meet the objectives and constraints.}}
	\label{fig_LLMs7}
   	\vspace{-1.2em}
\end{figure*}
\subsubsection{Scenario Description}
\par UAVs are extensively used for environmental monitoring and data \textcolor{b4}{collection. In such applications it is important to design} a flight \textcolor{b4}{route for a UAV that} encompasses all necessary monitoring points \textcolor{b4}{and to allocate} resources reasonably, thus revealing its executive capability and potential risks under extreme conditions. As shown in Fig. \ref{fig_LLMs7}, we consider a dynamic network consisting of a UAV, a UAV operator, an energy supply station, and $N$ necessary monitoring points. The fully charged UAV starts from the initial monitoring point \textcolor{b4}{$A$, and visits} and collects data from each monitoring point exactly once, where the UAV is charged once at the energy supply station and must be fully charged to continue the task. The energy consumption for the UAV \textcolor{b4}{includes the energy needed for both flying and data collection. It is} important to note \textcolor{b4}{that the energy} consumption of data collection can be substantial compared with energy consumption of flight, and its impact on overall energy consumption should not be ignored. \textcolor{b3}{Moreover, efficient data transmission depends on the quality of network coverage and the availability of communication resources such as bandwidth and transmission power. Poor network conditions can lead to higher energy consumption due to retransmissions and increased power requirements. Therefore, optimizing communication resource allocation based on data volume and network conditions is essential.} After the task is completed, the UAV returns to the starting monitoring point $A$. The optimization objective is to minimize the remaining energy of the UAV upon returning to the initial monitoring point $A$, and ensuring that the task is not interrupted due to energy depletion.

\begin{figure*}[!hbt] 
    \centering
    \subfigure[]{
        \begin{minipage}[t]{0.48\textwidth} % 调整这里的宽度比例
            \flushleft % 左对齐
            \includegraphics[width=\linewidth]{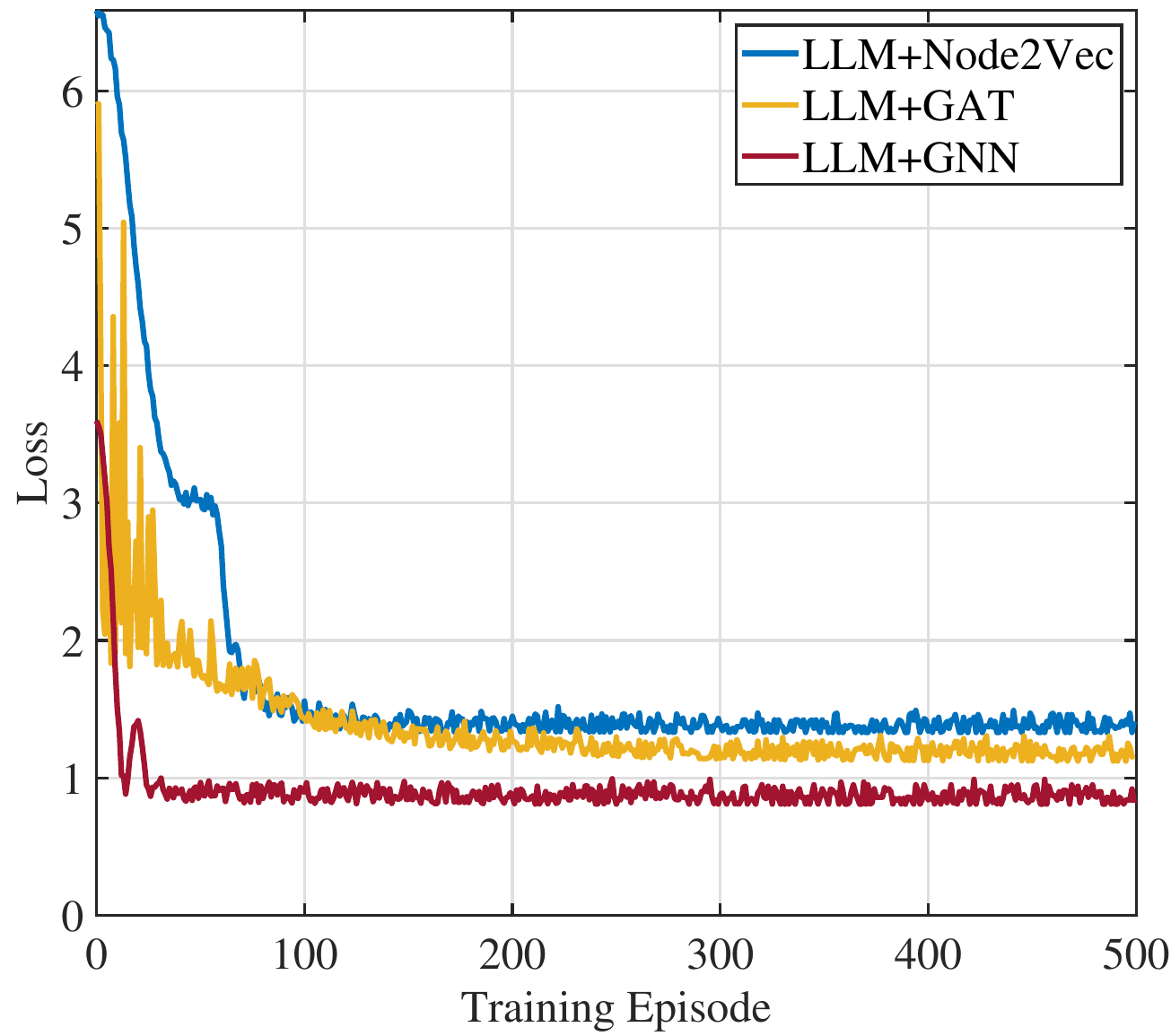}
        \end{minipage}
    }\hfill% <-- 使用 \hfill 自动填充适当的间距
    \subfigure[]{
        \begin{minipage}[t]{0.48\textwidth} % 调整这里的宽度比例
            \flushright % 右对齐
            \includegraphics[width=\linewidth]{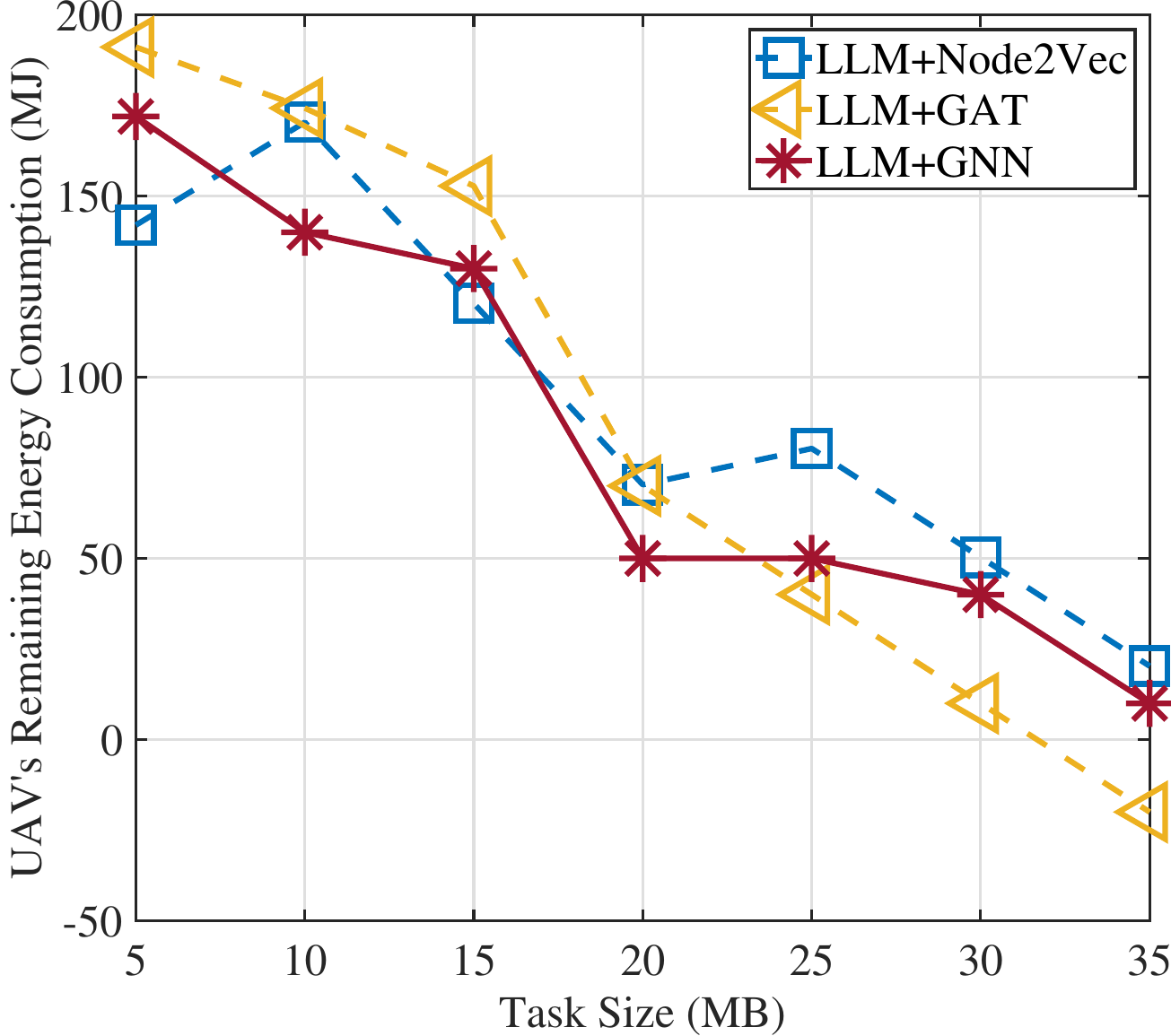}
        \end{minipage}
    }
    \caption{\textcolor{b4}{Experimental system performance results for UAV optimization. (a)} The difference \textcolor{b4}{between measurements and estimates. (b)} The impact of task size at monitoring points \textcolor{b4}{on the UAV's remaining} energy consumption. \textcolor{b3}{Note that the number of monitoring points is 6, \textcolor{b4}{in the narrative, $C$ is the charging station.} The amount of data that needs to be collected at each monitoring point \textcolor{b4}{is the task size (in MBs)}. \textcolor{b4}{The} UAV flies at an altitude of 100 $m$, and the area that the UAV can monitor is $400\times 400 \ \text{m}^2$}.}
    \label{fig_LLM+GNN}
    \vspace{-1.2em}
\end{figure*}

\subsubsection{Framework Workflow}
\par To leverage the graph-structured topology information of dynamic networks while handling the optimization objective, we
convert the considered scenario into a graph denoted by $G = (V, E)$, where $V$ denotes the set of
nodes (e.g., UAV or monitoring points), and $E$ denotes the set of edges (\textcolor{b4}{e.g., the UAV} collects data from $A$). In solving the UAV trajectory and \textcolor{b}{communication resource} allocation \textcolor{b4}{problem to satisfy the objectives}, we adopt a two-stage joint optimization strategy. Specifically, in the first stage, key features such \textcolor{b4}{as the starting} position of the UAV, distance between monitoring points, and the amount of data to be collected are extracted from the graph $G$. These features are then converted into \textcolor{b4}{text} and input into an LLM to generate the initial trajectory of the UAV. Subsequently, the trajectory information is combined with the original graph $G$ as input to the second stage through residual connections. Then, these combined \textcolor{b4}{data points are} further \textcolor{b4}{linearly} transformed and \textcolor{b4}{nonlinearly} mapped at the fully connected layer to generate higher-dimensional and richer feature representations, thus providing richer information \textcolor{b4}{to a GNN for} generating \textcolor{b}{communication resource allocation} strategies. During the process, a joint optimization loss function is designed to consider the outputs from both stages, \textcolor{b4}{enabling the GNN to optimize the UAV} trajectory and \textcolor{b}{communication resource} allocation by minimizing the loss function during training. This will ensure coordination between the two and achieve the optimization \textcolor{b4}{objectives}.

\subsubsection{Evaluation Results} Fig. \ref{fig_LLM+GNN}(a) presents \textcolor{b4}{the difference between measurements and estimates} of the proposed LLM+GNN and \textcolor{b4}{compares} it \textcolor{b}{with} those of LLM+Node2Vec, and LLM+GAT. The decreasing curve in the figure shows that the gap between \textcolor{b4}{the estimated and measured values} decreases as the number of training episodes increases. \textcolor{b4}{It is seen that} our proposed LLM+GNN outperforms LLM+Node2Vec and LLM+GAT. Therefore, the abovementioned results demonstrate the effectiveness of LLM+GNN in optimizing trajectory and \textcolor{b}{communication resource} allocation for UAV networks.

\par Moreover, Fig. \ref{fig_LLM+GNN}(b) shows the impact of task size at monitoring points \textcolor{b4}{on the UAV's remaining} energy consumption performance. As the amount of data collected (i.e., task size) at monitoring points increases, the remaining energy consumption of the UAV gradually decreases. This is because the UAV \textcolor{b4}{needs} to adjust its flight trajectory and allocate more resources to the monitoring points \textcolor{b4}{with larger tasks, which} results in increased energy consumption. This \textcolor{b4}{enables the UAV} to gather more information and execute environmental monitoring operations accurately. It is worth noting that the LLM+Node2Vec performs optimally when the task size is 5 \textcolor{b}{MB}. This is because Node2Vec could effectively capture both local and global structures in small-scale networks and generate efficient embeddings by optimizing the proximity of nodes, thereby extracting key information. Furthermore, when the task size increases to \textcolor{b}{25 MB and 30 MB}, the performance of LLM+GAT is slightly better than our proposed LLM+GNN. This can be attributed to the graph attention mechanism of GAT, which can dynamically allocate resources and priorities based on the importance of nodes, thereby optimizing energy consumption. When the task size further increases to 35 MB, the remaining energy of the UAV shows negative values, indicating that the task is interrupted due to energy exhaustion. This is because the attention mechanism of GAT may lead to the over-allocation of resources to certain important nodes, neglecting the needs of other nodes. Accordingly, our proposed LLM+GNN performs \textcolor{b4}{more stably and reliably under} various task sizes, which means that it is especially suitable \textcolor{b4}{for heavily loaded} and resource-limited UAV systems.

\section{Future Directions}
\par In this section, we will \textcolor{b4}{present some interesting future} directions of \textcolor{b}{LLM-enabled} graphs in dynamic networking.
\subsection{Adaptive Network Optimization}
\label{sec_Adaptive Network Optimization}
\par Adaptive network management is perhaps the most critical application of LLMs in dynamic networking. By leveraging LLMs to analyze real-time data from network graphs, networks can dynamically adjust to varying conditions. This capability allows for the optimization of traffic flow, real-time anomaly detection, and automated troubleshooting, ensuring that the network remains resilient and performs efficiently under different scenarios.

\subsection{Enhanced Security and Threat Detection}
\label{sec_Enhanced Security and Threat Detection}
\par Security is a paramount concern in dynamic networking, and LLMs may offer significant advancements in this area. By continuously monitoring and analyzing network graphs, LLMs could identify unusual patterns and potential threats. This proactive approach to threat \textcolor{b4}{detection would allow for} faster response times and more effective mitigation strategies, thereby protecting the network from malicious activities and ensuring the safety of sensitive data.

\subsection{Intelligent Network Automation}
\label{sec_Intelligent Network Automation}
\par Intelligent network automation driven by \textcolor{b4}{LLMs have the potential to significantly reduce} the complexity involved in managing modern networks. LLMs can interpret high-level network policies written in natural language and translate them into specific configurations. This automation minimizes the need for manual intervention, reduces human error, and speeds up the deployment of network services, making network management more efficient and scalable.

\section{Conclusion}
\par In this article, \textcolor{b4}{we have explored} the integration of LLMs and graphs in dynamic networks. Specially, we first \textcolor{b4}{introduced} the related \textcolor{b4}{concepts}, key technologies, and applications of \textcolor{b}{LLM-enabled} graphs. Following this, we \textcolor{b}{presented} \textcolor{b}{LLM-enabled} graphs and their applications in dynamic networks from the perspective of LLMs as different roles such as \textcolor{b4}{predictors, encoders and aligners.} Subsequently, we \textcolor{b}{proposed} a novel framework of \textcolor{b}{LLM-enabled} graphs for networking optimization, and then \textcolor{b4}{conducted} a case study on UAV networking, focusing on \textcolor{b4}{optimizing a UAV's trajectory} and \textcolor{b}{communication resource} allocation to validate the effectiveness of the proposed \textcolor{b}{LLM-enabled} graphs framework. Finally, we \textcolor{b4}{discussed some future directions for the study of LLM-enabled} graphs in dynamic networks.

\ifCLASSOPTIONcaptionsoff
\newpage
\fi

\bibliographystyle{IEEEtran}
\bibliography{references.bib}

\end{document}